\documentstyle[psfig]{article} 
\def\etal   {{\it et al.}~}
\def\chandra {{\it Chandra}~}

\newcommand{\lya}{\ifmmode {\rm Ly\,}\alpha \, \else Ly\,$\alpha$\,\fi}

\newcommand{\kms}{\ifmmode {\rm km\,s}^{-1} \else km\,s$^{-1}$\fi}
\newcommand{\ergsA}{\ifmmode {\rm ergs\,cm}^{-2}\,{\rm s}^{-1}\,{\rm \AA }^{-1}
\else ergs\,cm$^{-2}$\,s$^{-1}$\,\AA $^{-1}$\fi}

\begin{document}

\centerline{\Large \bf
Warm Absorbers in Active Galactic Nuclei}  

\bigskip
\centerline{\bf Smita Mathur}
\centerline {The Ohio State University}

\bigskip
\section*{Abstract}

Observations of warm absorbers provided new ways to study the nuclear
environments of AGNs. I discuss basic properties of warm absorbers and
early developments with ROSAT, ASCA and HST observations. I briefly
touch upon recent advances made with \chandra and FUSE.

\medskip 
{\it Key-words}{ : quasars: absorption lines-- ultraviolet: galaxies--
X-rays: galaxies}

\section{Introduction: Warm absorbers with ROSAT and ASCA}

What is a X-ray warm absorber? ROSAT observations of some Seyfert
galaxies showed that a simple power-law does not describe the data
adequately. Often strong residuals were observed around 0.8 keV
(Figure 1). These residuals were interpreted as absorption edges due
to highly ionized oxygen OVII/OVIII and so the absorber was called
warm or ionized. Theoretically, such absorption features are expected
when continuum flux passes through a slab of ionized matter (Halpern,
1982; Reynolds \& Fabian 1995). The exact signature of a warm absorber
depends upon its column density, shape of the ionizing continuum and
ionization state (usually parameterized in terms of ionization
parameter which is a dimentionless ratio of photon to electron
density). Figure 2 shows the dependence of warm absorber signatures on
column density and ionization parameter U.

\vspace{-0.50cm}
\begin{figure}[h]
\begin{center}
\psfig{figure=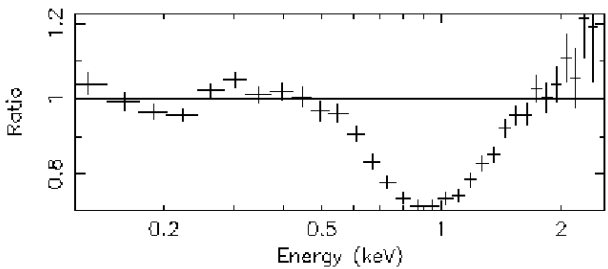,height=4.0cm,angle=-90}
\end{center}
\vspace{-0.85cm}
\caption{\footnotesize Ratio of ROSAT data of NGC 3516 to a simple
power-law model fit. The strong residuals at $\sim$ 0.9 keV are the
signature of a warm absorber. (From Mathur \etal 1997)}
\label{fig1}
\end{figure}

Signatures of ionized gas from nuclear regions of AGNs were known to
be present in the UV band as well, from IUE (e.g. Ulrich 1988) and HST
(e.g. Bahcall \etal 1993) studies. However, the physical properties of
the absorber were not known as absorption lines of only a few elements
were observed, typically CIV, NV, \lya and in some cases
OVI. Quasi-simultanous observations of quasar 3C351 with ROSAT and HST
helped solve the problem. ROSAT data of 3C351 detected absorption
edges due to OVII/OVIII and HST observations showed OVI absoprtion
lines. Through detailed photoionization modeling, Mathur
\etal (1994) showed that both X-ray and UV signatures arise from the
same absorbing material.
%\vspace{-2.0cm}
\begin{figure}[h]
\begin{center}
\hbox{
\psfig{figure=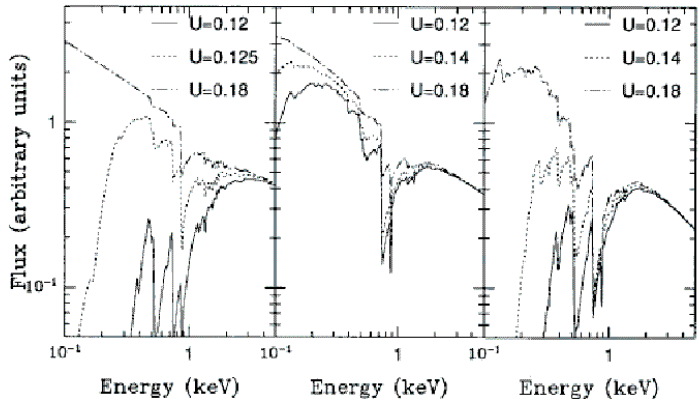,height=6.0cm,angle=-90}
}
\end{center}
\vspace{-0.75cm}
\caption{\footnotesize Behavior of a warm absorber with changes in
column density N$_H$, plower-law continuum slope $\alpha$ and
ionization parameter U. Left: $\alpha=0.5$, N$_H=3.5 \times 10^{22}$
cm$^{-2}$; Middle: $\alpha=0.6$, N$_H=1.2 \times 10^{22}$ cm$^{-2}$;
Right: $\alpha=0.0$, N$_H=1.2 \times 10^{22}$ cm${-2}$ (from Fiore
\etal 1993).}
\label{fig1}
\end{figure}

\begin{figure}[h]
\begin{center}
\psfig{figure=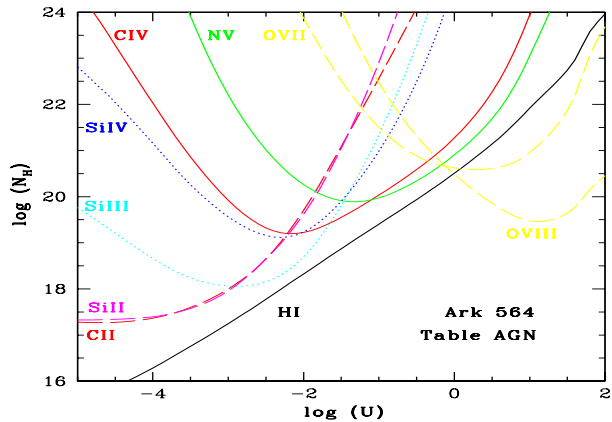,height=7.0cm,width=10.0cm}
\end{center}
\vspace{-1.0cm}
\caption{\footnotesize Constarints on parameter space using
photoionization models and observed line strengths. Dotted lines
correspond to observed values of column density of individual ions;
dashed lines mark upper limits and solid lines mark lower limits. The
allowed range is then: $-1.5<\log U<-1$, and $20<\log N_H<21.5$ (from
Romano \etal 2002). }
\label{fig1}
\end{figure}

Since the discovery of a unified XUV absorber in 3C351, many
observations of many AGNs showed one to one correspondence between
absorption in X-rays and in UV (Crenshaw 1997, Monier \etal 2001 and
references there in). Because the UV and X-tay absorbers were found to
be one and the same, constraints from both the datasets could be
combined to determine the physical conditions in the absorber (see,
e.g. table 3 in Mathur \etal 1995). A previously not recognized
component of nuclear material was discovered: high column density,
highly ionized, outflowing matter situated in or around the broad
emission line region. This lead to important realization that the mass
outflow rate in many cases is comparable to the accretion rate, and
that the outflow carries a significant amount of kinetic
energy. Understanding of the physical conditions, kinematics and
geometry of the absobing matter lead to one of the first comprehensive
models of AGN structure (Elvis 2000).

Later observations with ASCA could resolve the OVII edge from OVIII
providing better constraints on the physical state of the absorber
(George \etal 2000).  Photoionization models (figure 3) of X/UV
absorbers also led to the prediction that in addition to the edges,
resonance absorption lines due to highly ionized elements should also
be present (Nicastro \etal 1999). These could only be observed with
high resolution spectroscopy. Complexity, as commonly seen in UV
absorption systems is also expected in X-ray warm absorbers (see
Mathur 1997 and references there in).

\section{\chandra Observations of Warm Absorbers}

High resolution spectroscopy of AGNs known to have warm absorbers, was
performed in the first two years of \chandra. Here I will discuss only
one example and refer readers to Mathur (2001) for a review on
\chandra results on AGNs. 

\noindent
{\it NGC 5548}: This well studies AGN was observed with
LETG/HRC. Strong, narrow absorption lines from highly ionized species
are clearly present in the spectrum (Kaastra \etal 2000). The widths
of the X-ray lines are consistent with the UV lines. The average
blueshift of the X-ray absorption lines was found to be somewhat
smaller, but comparable to the UV lines. However, further relation
between the X-ray and UV absorbers remained an open issue because of
the low S/N and lower resolution of the X-ray spectrum. (The spectral
resolution of LETG is $\sim$300--1000 compared to $\sim$10,000 of
HST/STIS in medium resolution.)\\

The absorption lines/ edges discussed above are observed when there is
absorbing gas along the line of sight. The same plasma, when viewed
from another angle should exhibit emission lines. In addition to the
resonance lines, seen in absorption, forbidden and intercombination
lines should also be present in emission (see also Krolik \& Kriss
1995). Such emission lines are best studied when the bright nuclear
continuum is suppressed. Naturally, the first \chandra observations to study
emission lines from AGNs were of highly absorbed Seyfert
galaxies. Below I review one such example. 

\noindent
{\it Mrk 3}: \chandra HETG/ACIS-S observation of this Seyfert 2 galaxy
is reported by Sako \etal (2000). A number of lines from a variety of
elements were detected in the spectrum. Resonance lines and Fe-L
lines, characteristic of photoionized plasma are strong. The OVII
triplet, useful for plasma diagnostics, is also detected. Sako \etal
conclude that the emission line plasma is clearly photoionized, with
practically no contribution from collisionally ionized gas. As such,
the plasma characteristics are consistent with a warm absorber seen in
emission.

\noindent
All the new observations of warm absorbers have shown
consistency with the model predictions based on a common
origin of the X-ray and UV absorbing gas.

\section{UV Studies of Warm Absorbers}

UV studies of ionized gas in the nuclear regions of AGNs are done by
observing resonance lines of highly ionized ions of mainly CIV and NV,
as they were easily accessible first with IUE and then with HST. The
higher ionization OVI line was not accessible by these missions for
low redshift Seyfert galaxies. OVI observations, however, are very
important because together with OVII/OVIII in the X-ray band they
offer best constarints on the ionization state of the absorber. This
barrier was overcome with the launch of FUSE. Seyfert galaxies known
to harbor X-ray warm absorbers indeed showed associated OVI absorption
in FUSE spectra (see Kriss 2001). Figure 4 shows FUSE spectrum of a
narrow line Seyfert 1 galaxy Akn 564. Strong OVI absorption is clearly
present. OVII and OVIII absorption lines are present in the high
resolution \chandra spectrum (Mastumoto, Leighly \& Marshall 2001).

\begin{figure}[h]
\begin{center}
\psfig{figure=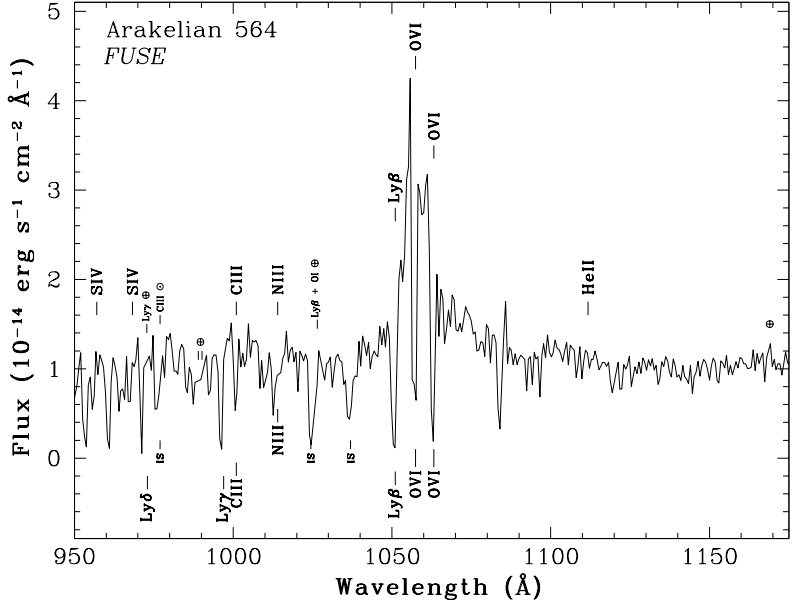,height=9.0cm,angle=-90}
\end{center}
\vspace{-1.0cm}
\caption{\footnotesize FUSE spectrum of Akn 564 showing strong absorption from OVI and HI Lyman series (from Romano \etal 2002).}
\label{fig1}
\end{figure}

\section{Conclusions}

High resolution observations of warm absorbers have resulted in
beautiful spectra. However, they have not yet given us additional
insight into the properties/ kinematics of the absorber. This is
because most of the absorption line spectra have low signal to noise
ratio (S/N): the measured absorption line equivalent widths are
uncertain to $\pm 40\%$, larger than those from sensitive ASCA 
observations. High resolution \chandra spectra should be viewed more
as a proof of concept than diagnostic tools. Moreover, most of the
modeling efforts are preliminary. So interpretations should be made
with caution, especially if the results are unexpected. Only when we
obtain well constrained parameters can we hope to build and test
models of AGN structure and physics. One such effort is lead by Ian
George and Mike Crenshaw in obtaining multiwavelength, high S/N
spectra of a nearby Seyfert galaxy with a warm absorber. Our team has
observed NGC 3783 with HETG/ACIS-S for 900 ksec, with HST/STIS for 34
orbits and also with FUSE and XTE. These observations have resulted in
one of the best high resolution spectra of AGNs (figure 5). Modeling
efforts are under way (Netzer \etal, in preparation). Within next year
or so, we will know whether or not our view of the nuclear region of
AGNs gets altered dramatically.

Because of space limitations, I could not cover the subject
fully. Interested readers are referred to conference proceedings:
``Mass Ejection from AGN'' and ``Mass Outflow in Active Galactic
Nuclei: New Perspectives". I thank the organizers of the meeting for
inviting me to give a review talk on warm absorbers in AGNs. This has
been a very nice meeting and it is always wonderful to be back home.

\begin{figure}[h]
\begin{center}
\psfig{figure=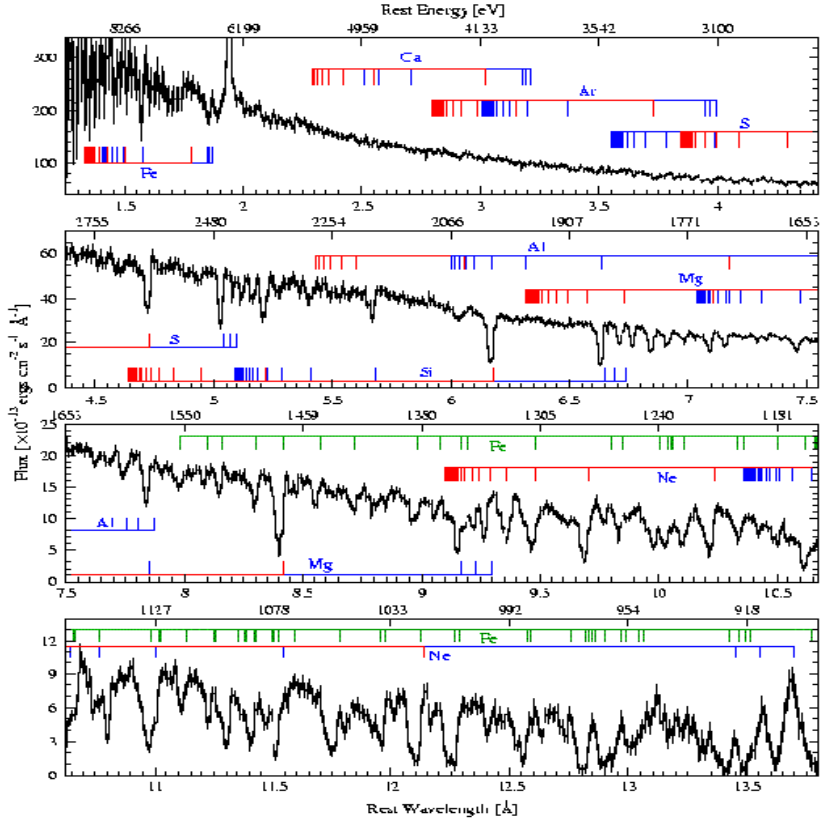,height=12.0cm,width=12.0cm}
\end{center}
\vspace{-1.0cm}
\caption{\footnotesize 
A long look HETG/ACIS-S spectrum of NGC 3783. Note the high S/N in this low wavelength region (From Kaspi \etal 2002).}
\label{fig1}
\end{figure}

\section*{References}

%Please follow the ApJ style for the references
%Separate the references by a blank line

\noindent
Bahcall, J.N. \etal 1993, ApJS, 87, 1

\noindent
Crenshaw, M. 1997 in ``Mass Ejection from AGN'', Ed: N. Arav, I., Shlos
man, \& R. Weymann [ASP Conf. Series Vol. 128]

\noindent
Elvis, M. 2000, ApJ, 545, 63

\noindent
Fiore, F., Elvis, M., Mathur, S., Wilkes, B., \& McDowell J. 1993, ApJ, 415, 129 

\noindent
George, I. \etal 2000, ApJ, 531, 52

\noindent
Halpern, J. 1982, Ph.D. Thesis, Harvard University. 

\noindent
Kaastra, J. \etal 2000, A\&A Lett., 354, 83

\noindent
Kaspi, S. \etal 2002, ApJ, submitted

\noindent
Kriss, G. 2001, in ``Probing the Physics of Active Galactic Nuclei'', Ed: B.M. Peterson, R.W. Pogge, and R.S. Polidan. [San Francisco: ASP (vol 224)]

\noindent
Krolik, J, \& Kriss, G. 1995, ApJ, 447, 512

\noindent
Mastumoto, C., Leighly, K. \& Marshall, H. 2001, in ``X-ray Emission from Accretion onto Black Holes'', Eds: T. Yaqoob.

\noindent
Mathur, S., Wilkes, B., Elvis, M. \& Fiore, F. 1994, ApJ, 434, 493 

\noindent
Mathur, S., Elvis, M. \& Wilkes, B. 1995 ApJ, 452, 230

\noindent
Mathur, S. 1997 in ``Mass Ejection from AGN'', Ed: N. Arav, I., Shlos
man, \& R. Weymann [ASP Conf. Series Vol. 128]

\noindent
Mathur, S. 2001, in ``The High Energy Universe at Sharp Focus: Chandra Symposium", Ed: E. Schlegel and S. Vrtilek [ASP Conf. Series]

\noindent
Mathur, S., Wilkes, B., \& Aldcroft, T. 1997, ApJ, 478, 182 

\noindent
Monier, E., Mathur, S., Wilkes, B., \& Elvis, M. 2001, ApJ, 559, 675

\noindent
Nicastro, F., Fiore, F. \& Matt, G. 1999, ApJ, 517, 108

\noindent
Reynolds, C.S. \& Fabian, A.C. 1995, MNRAS, 273, 1167 

\noindent
Romano, P., Mathur, S., Pogge, R. \& Peterson, B.M. 2002, in preparation

\noindent
Sako, M., Kahn, S.M., Paerels, F., Liedahl, D. 2000, ApJL, 543, 115

\noindent
Ulrich, M.H. 1988, MNRAS, 239, 121

\end{document}